# On the applicability of Kramers-Kronig dispersion relations to guided and surface waves

Victor V. Krylov

*Department of Aeronautical and Automotive Engineering, Loughborough University, Loughborough, Leicestershire LE11 3TU, UK*

**ABSTRACT**

In unbounded media, the acoustic attenuation as function of frequency is related to the frequency-dependent sound velocity (dispersion) via Kramers-Kronig dispersion relations. These relations are fundamentally important for better understanding of the nature of attenuation and dispersion and as a tool in physical acoustics measurements, where they can be used for control purposes. However, physical acoustic measurements are frequently carried out not in unbounded media, but in acoustic waveguides, e.g. inside liquid-filled pipes. Surface acoustic waves are also often used for physical acoustics measurements. In the present work, the applicability of Kramers-Kronig relations to guided and surface waves is investigated using the approach based on the theory of functions of complex variables. It is demonstrated that Kramers-Kronig relations have limited applicability to guided and surface waves. In particular, they are not applicable to waves propagating in waveguides characterised by the possibility of wave energy leakage from the waveguides into the surrounding medium. For waveguides without leakages, e.g. those formed by rigid walls, Kramers-Kronig relations remain valid for both ideal and viscous liquids. In the former case, Kramers-Kronig relations express the exponential decay of non-propagating (evanescent) higher-order acoustic modes below the cut-off frequencies via the dispersion of the same modes above the cut-off frequencies. Examples of numerical calculations of wave dispersion and attenuation using Kramers-Kronig relations, where applicable, are presented for unbounded media and for waveguides formed by rigid walls.

## 1. Introduction

It is a matter of common knowledge that in unbounded media the acoustic attenuation as function of frequency is linked to the frequency-dependent sound velocity (dispersion) via Kramers-Kronig dispersion relations, or simply Kramers-Kronig relations. These relations have been named after Kramers [1] and Kronig [2] who established them independently for attenuation and dispersion of electromagnetic waves (in the field of Optics). In Acoustics, these relations have been first described by Ginzburg [3] in respect of sound attenuation and dispersion. This publication was followed by a number of papers published by different authors who examined applications of Kramers-Kronig relations to various problems linked to investigations of acoustic wave attenuation and dispersion in unbounded media (see e.g. [4-13]). Note that Kramers-Kronig relations are the consequence of causality and linearity of the medium [14]. This is why they are very general and fundamentally important for better understanding the nature of wave attenuation and dispersion and as a useful mathematical tool in physical acoustics measurements, where they can be used for verification purposes.

It should be noted that physical acoustics measurements are often carried out not in unbounded media, but in acoustic waveguides, e.g. inside liquid-filled pipes. Surface acoustic waves are also used frequently for physical acoustics measurements. As a rule, guided or surface waves are dispersive even in the idealised situations of absence of material losses in the medium, which seems to be in contradiction with Kramers-Kronig relations requiring the presence of wave attenuation in such cases. This raises the question of applicability of Kramers-Kronig relations to guided and surface waves.

The problem of applicability of Kramers-Kronig relations to acoustic waveguides has been first examined by the present author [15] using the inversion theorem of the theory of functions of complex variables. It has been shown that Kramers-Kronig relations have limited applicability to guided and surface waves. In particular, they are not applicable to waves propagating in waveguides characterised by the possibility of wave energy leakage from the waveguides into the surrounding medium. For waveguides without leakages, e.g. liquid-filled pipes with rigid walls, Kramers-Kronig relations may remain valid for both ideal and viscous liquids. In the former case, Kramers-Kronig relations express the exponential decay of non-propagating (evanescent) higher-order acoustic modes below the cut-off frequencies via the dispersion of the same modes above the cut-off frequencies. Note that the latter result has been rediscovered in the later paper by Haakestad et al. [16]. Apparently, the authors of Ref. [16] were unaware of the earlier work [15], and they used a different approach in their analysis.

In the present paper, the applicability of Kramers-Kronig relations to guided and surface acoustic waves is further investigated using the approach developed in Ref. [15]. It should be noted that the term "acoustic" is used in the physics community for all mechanical waves caused by elasticity of the media, which includes sound waves in gases and liquids as well as elastic waves in solids. In the mechanical engineering community, the term "acoustic" is used only for "classical" air-borne or underwater sound, whereas all numerous types of elastic waves in solids are considered as "vibrations". In this paper, the term "acoustic" is used in the same way as it is used in the physics community, i.e. for all elastic waves in gases (liquids) and in solids.

Examples of numerical calculations of wave dispersion and attenuation using Kramers-Kronig relations are presented for two cases where they are applicable: for unbounded media with relaxation and for guided waves in ideal waveguides with rigid walls. Comparison of these numerical results with the direct analytical calculations of sound attenuation and dispersion, which are possible in these cases, illustrate the importance of Kramers-Kronig relations for both general wave theory and practical applications. Part of the results described in this paper have been presented at the conference 'Acoustics 2014' (Birmingham, UK) and published in the conference proceedings [17].

## 2. Kramers-Kronig relations for waves in unbounded media

It is instructive to start a discussion of Kramers-Kronig relations with a common case of sound propagation in unbounded media. It is convenient to apply Kramers-Kronig relations to a complex coefficient of refraction $n(\omega) = c(\infty)/c(\omega)$, where $c(\infty)$ is the sound velocity in the medium at an infinitely large frequency, and $c(\omega)$ is the sound velocity in the same medium at arbitrary circular frequency $\omega$. The velocity $c(\omega)$ is generally assumed to be complex in order to take account of dissipation.

Kramers-Kronig dispersion relations can be written in the form (see e.g. [14])

$$\mathrm{Re}\, n(\omega) - 1 = \frac{1}{\pi} V.P. \int_{-\infty}^{\infty} \frac{\mathrm{Im}\, n(\omega')}{\omega' - \omega} d\omega', \qquad (1)$$

$$\mathrm{Im}\, n(\omega) = -\frac{1}{\pi} V.P. \int_{-\infty}^{\infty} \frac{\mathrm{Re}\, n(\omega') - 1}{\omega' - \omega} d\omega'. \qquad (2)$$

Here *Re n(ω)* and *Im n(ω)* are real and imaginary parts of *n(ω)* respectively, and symbol 'V.P.' indicates that the integrals in (1) and (2) are understood in the sense of the Cauchy principal value ('valeur principale'). Note that mathematical derivation of the relations (1) and (2) is based on the Cauchy theorem and on the analyticity of the function *n(ω)* in the upper or lower half spaces of the complex variable *ω*, which in turn usually follows from the causality principle [14], which means that the reaction of the medium cannot take place before the action. Also, a 'good' behaviour of *n(ω)* for large values of *ω* is assumed.

In many cases functions *n(ω)* have some symmetry in respect of frequency, namely *n(-ω) = n\*(ω)*. In such cases Kramers-Kronig dispersion relations (1) and (2) can be rewritten in the form of the integrals over positive frequencies only:

$$\mathrm{Re}\, n(\omega) - 1 = \frac{2}{\pi} V.P. \int_{0}^{\infty} \frac{\omega' \mathrm{Im}\, n(\omega')}{(\omega')^2 - \omega^2} d\omega', \qquad (3)$$

$$\mathrm{Im}\, n(\omega) = -\frac{2}{\pi} V.P. \int_{0}^{\infty} \frac{\omega [\mathrm{Re}\, n(\omega') - 1]}{(\omega')^2 - \omega^2} d\omega'. \qquad (4)$$

As an example, let us consider application of Kramers-Kronig relations to the case of a medium with relaxation, for which a full complex function for the wavenumber *k(ω)* is well known (see e.g. [18]):

$$k(\omega) = \frac{\omega}{c(0)} \left[ 1 - \frac{p}{2} \frac{\omega^2 \tau^2}{1 + \omega^2 \tau^2} + i \frac{p}{2} \frac{\omega \tau}{1 + \omega^2 \tau^2} \right]. \qquad (5)$$

Here $\tau$ is the time of relaxation for a given medium, and $p = [(c(\infty))^2 - (c(0))^2]/(c(0))^2$. Expressing the coefficient of refraction $n(\omega) = c(\infty)/c(\omega)$ from (5), one can obtain

$$n(\omega) = \frac{1}{1 - p/2} \left[ 1 - \frac{p}{2} \frac{\omega^2 \tau^2}{1 + \omega^2 \tau^2} + i \frac{p}{2} \frac{\omega \tau}{1 + \omega^2 \tau^2} \right]. \qquad (6)$$

Plot of the real part of *n(ω)* defined analytically by (6) is shown in Fig. 1. Calculations have been carried out for *p = 0.089* and *τ = 0.0003* s.

Let us now calculate the imaginary part of *n(ω)* numerically using Kramers-Kronig relations. Substituting the real part of *n(ω)* from (6) into (4) and performing numerical integration, one can obtain the imaginary part, *Im n(ω)*. The results of the calculations are shown in Fig. 2 by a dashed curve. For comparison, the imaginary part of *n(ω)* defined by Equation (6) is also displayed in Fig. 2, by a solid curve. As expected, both these curves perfectly coincide with each other, which illustrates that Kramers-Kronig relations work well in this case.

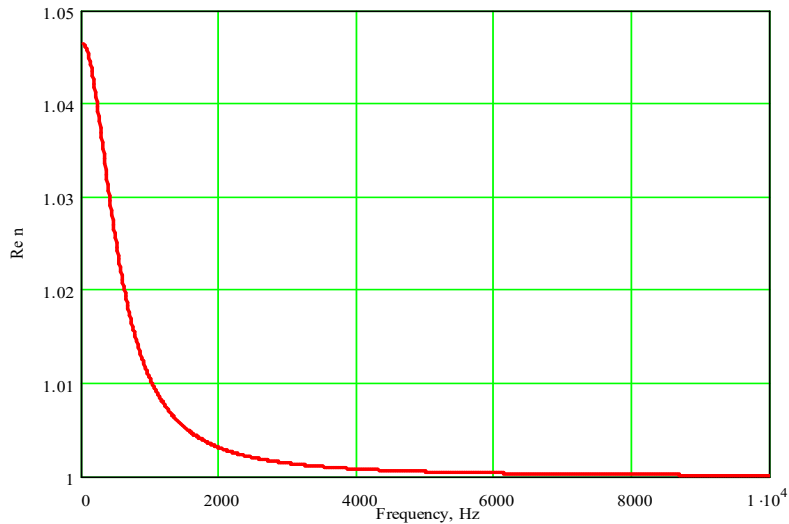

Fig. 1  Real part of the refraction coefficient *n(ω)* for a medium with relaxation defined by Equation (6).

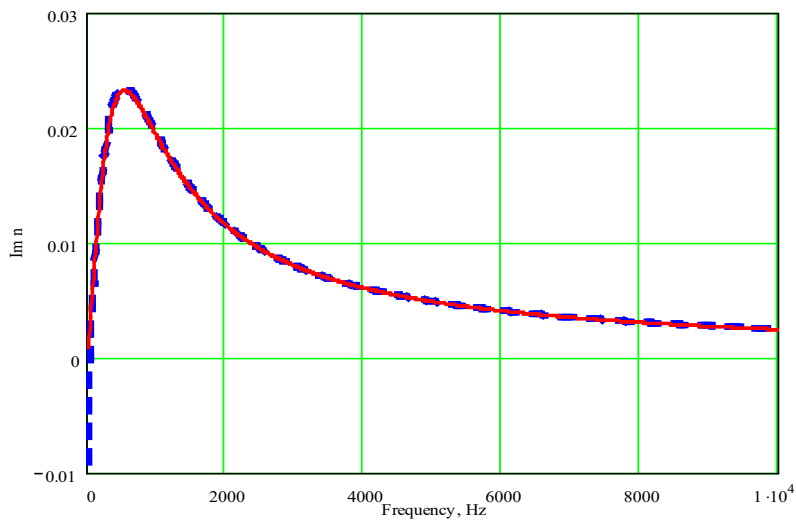

Fig. 2  Imaginary part of the refraction coefficient *n(ω)* for a medium with relaxation defined by Equation (6) (solid curve) and calculated using Kramers-Kronig relation (4) (dashed curve).

## 3. Kramers-Kronig relations for guided waves

### 3.1 General remarks

In the case of guided waves, the behavior of the complex refraction index $n(\omega)$ is determined not only by the material properties of the medium, like in the case of unbounded media considered above, but also by the dispersion equations defined for specific waveguiding systems, e.g. elastic layers, liquid-filled pipes, etc. This makes the behavior of the functions $n(\omega)$ for guided modes much more complicated. As a result of this, their analyticity can be violated, causing the breakdown of Kramers-Kronig relations.

Some general conclusions for such waveguiding systems have been obtained in Ref. [15] using the inversion theorem of the theory of functions of complex variables. In particular, it has been shown that one of the common reasons for violation of analyticity can be the appearance of branch points on the complex plane $\omega$ that are associated with some physical features of the waveguides.

According to the inversion theorem of the functions theory [19], functions $n(\omega)$ may have a branch point of order $j$ at the point $\alpha \neq \infty$ if the derivative of the inverse function $d\omega/dn$ has a zero of order $j$ or a pole of order $j+2$ at the point $n(\alpha)$. Since zeros of the function $d\omega/dn$ coincide with zeros of the function $d\omega/dk$ describing a complex group velocity (here $k = \omega/c(\omega)$ is the wavenumber), important conclusions about the positions of the branch points of the function $n(\omega)$ can be made from the positions of zero group velocity.

Generally speaking, branch points can be located at arbitrary positions on the complex plane $\omega$. In particular, they can be located away from the real axis $\omega$. In such cases, contour integration used for derivation of Kramers-Kronig relations would require adding integrals over the branch cuts, which means that Kramers-Kronig relations in such situations would become invalid.

An important exception are the cases where branch points of the functions $n(\omega)$ for normal waves are located on the real axis $\omega$. This means that the group velocities of such normal modes are equal to zero at critical (cut off) frequencies, which can occur for waveguides without possible leakage of energy from waveguides [20]. In such cases Kramers-Kronig relations may remain valid.

### 3.2 Acoustic waveguides formed by two rigid walls

As an example of waveguides without energy leakage, let us consider an acoustic waveguide made up by two rigid walls separated by the distance $h$ and filled with air (see e.g. Ref. [20]). The dispersion equation for such a waveguide is well known, and in terms of $n(\omega)$ it can be written in the form

$$n(\omega) = \left(1 - \frac{c^2(\infty)\pi^2 m^2}{\omega^2 h^2}\right)^{1/2}, \qquad (7)$$

where $m = 0, 1, 2, \ldots$. As it can be seen from (7), the function $n(\omega)$ has branch points at the cut off frequencies $\omega_m = \pm c(\infty)\pi m/h$ and a simple pole at $\omega = 0$.

Kramers-Kronig relations for this case can be written in the form [15]:

$$\operatorname{Re} n(\omega) - 1 = \frac{2}{\pi} V.P. \int_0^\infty \frac{\omega' \operatorname{Im} n(\omega')}{(\omega')^2 - \omega^2} d\omega', \quad (8)$$

$$\operatorname{Im} n(\omega) = -\frac{2}{\pi} V.P. \int_0^\infty \frac{\omega [\operatorname{Re} n(\omega') - 1]}{(\omega')^2 - \omega^2} d\omega' + \frac{c(\infty) \pi m}{h \omega}. \quad (9)$$

The only difference of these relations from the relations (3) and (4) is the presence of the last term in (9) caused by the pole at $\omega = 0$. As was first mentioned in Ref. [15], if the material attenuation in (7) is neglected, the dispersion relations (8) and (9) describe the relationship between the phase velocities of the propagating normal modes above the cut off frequencies (for $m \neq 0$) and the attenuation decrements of the non-propagating (evanescent) waves below these cut off frequencies (the case of $m = 0$ corresponding to the lowest order mode propagating without dispersion, $n(\omega) = 1$, is trivial and not considered here).

Numerical calculations carried out for the first mode ($m = 1$) of a waveguide formed by two parallel rigid walls and filled with air illustrate the above points. Calculations have been carried out for $h = 0.03$ m and $c(\infty) = 340$ m/s. Function $Re\ n(\omega)$ that has been calculated directly from the analytical expression (7) is shown in Fig. 3.

The results for $Im\ n(\omega)$ obtained from the numerical integration using the Kramers-Kronig dispersion relation (9) are shown in Fig. 4 (dashed curve) together with the results directly calculated from the analytical expression (7) below the cut off frequency (solid curve). As one can see, the solid and dashed curves in Fig. 4 are almost indistinguishable, which confirms validity of Kramers-Kronig relations in this case.

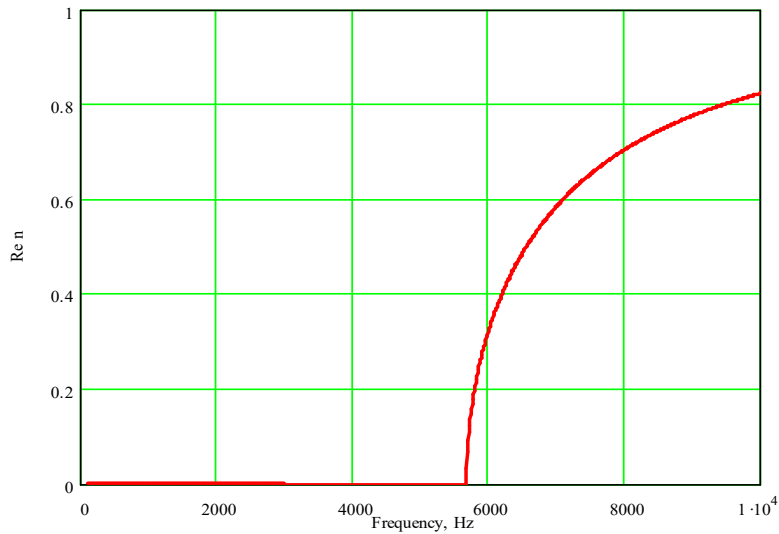

Fig. 3 Real part of the refraction coefficient $n(\omega)$ for the first mode ($m = 1$) of a waveguide with rigid walls defined by Equation (7).

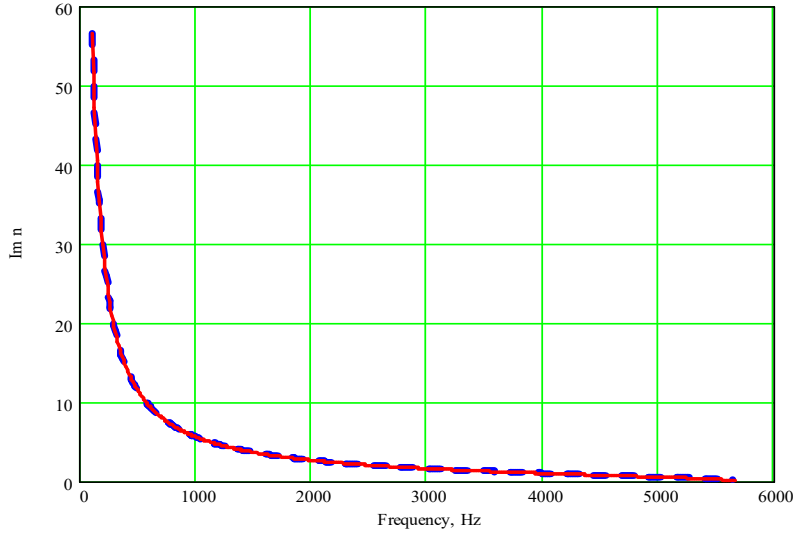

Fig. 4 Imaginary part of the refraction coefficient $n(\omega)$ for the first mode ($m = 1$) of a waveguide with rigid walls defined by Equation (7) (solid curve) and calculated using Kramers-Kronig relation (9) (dashed curve).

Note that Kramers-Kronig relations remain valid for waveguides with rigid walls also in the case of taking into account realistic material losses (e.g. if a waveguide is filled with a medium with relaxation). This is important for verification of physical acoustic measurements of sound attenuation and velocity in different materials and substances. If waveguide modes with $m \neq 0$ are used, Kramers-Kronig relations connect the combined dispersion due to relaxation and waveguide propagation with the combined attenuation of propagating and evanescent waves.

### 3.3 Open acoustic waveguides

Let us now discuss open waveguides (e. g. acoustic waveguides formed by horizontal layers with different sound velocities in the ocean or atmosphere [21]) and consider the case when wave energy leakage from such waveguides is not possible. Typical examples of such waveguides are infinite linear or parabolic layers that can be considered as approximate models for real waveguides. Such layers always have ray turning points, which means that there is no escape for the wave energy, so that there are no branch points of functions $n(\omega)$ outside the real frequency axis. However, the behavior of complex functions $n(\omega)$ at large $\omega$ for some of such waveguides still can be problematic. For example, they can depend on $\omega$ exponentially, in which case Kramers-Kronig relations may break down due to the fact that integrals over semi-infinite circles in the Cauchy contour integrals may not be equal to zero. The behavior of $n(\omega)$ at large $\omega$ should be investigated for each specific open waveguide before applying Kramers-Kronig relations.

In the case of open waveguides for which the possibilities of energy leakage can exist at some frequency ranges, the analytical properties of the complex refraction indexes of normal modes $n(\omega)$ can be broken down due to the appearance of branch points on the complex plane $\omega$ located away from the real axis $\omega$. For this reason, Kramers-Kronig relations for such waveguides generally become invalid. Even though branch cut integrals arising in the process of contour integration could be added in this case to obtain modified Kramers-Kronig relations, this would not make any practical sense as the location of branch points is specific for every particular waveguide, and it is generally unknown. Examples of open waveguides with the potential possibilities of energy leakage include typical ocean acoustic waveguides formed by horizontal layers with changing sound velocities [21].

### 3.4 Guided elastic waves in solids

Among the simplest guiding structures for elastic waves in solids are plates of constant thickness $h$. For the following brief discussion, we will assume that plates are placed horizontally. It is well known that such plates can support guided shear waves of horizontal polarization (SH modes) and guided waves polarized in the sagittal plane (Lamb waves). It should be noted that plates are not open waveguides for such guided waves.

Propagation of SH modes in plates does not differ in principle from sound wave propagation in classical acoustic waveguides formed by two parallel rigid walls considered in Section 3.2. There are no leakages of wave energy, so that Kramers-Kronig relations are applicable.

However, Kramers-Kronig dispersion relations are generally not applicable to Lamb waves propagating in elastic plates because, as it follows from the complex transcendental dispersion equations of Lamb modes, the functions $n(\omega)$ do not behave well at very large frequencies. These include the lowest order symmetric and antisymmetric Lamb modes, also called symmetric plate waves and flexural (bending) waves respectively, even though the velocities of both these mode tend to Rayleigh wave velocity at high frequencies.

Like in the case of acoustic waves in ocean, there are many examples of seismic open waveguides formed by layered ground for elastic waves of different polarisations [22], with the potential possibilities of energy leakage. In all such cases Kramers-Kronig relations are generally not applicable.

## 4. Kramers-Kronig relations for surface waves

### 4.1 Some general remarks

From the point of view of applicability of Kramers-Kronig relations, the case of surface waves is similar to the above-mentioned case of open waveguides. If the possibility of energy leakage exists for a particular type of surface waves, this is usually reflected in the appearance of branch points of the complex function $n(\omega)$ that are not located on the real axis $\omega$. Like in the case of open waveguides, Kramers-Kronig relations are generally not applicable to such surface waves.

### 4.2 Love waves

One of the typical examples of surface waves in solids are Love waves (see e.g. Ref. [23]), which are shear-horizontal elastic waves propagating in an elastic half space covered by an elastic layer with lower shear wave velocity. In their mathematical description, these waves are very similar to guided modes of open underwater acoustic waveguides discussed in the previous section. Energy leakage for such waves is possible, and therefore Kramers-Kronig relations are generally not applicable.

### 4.3 Rayleigh-type waves

Another example of surface waves are Rayleigh-type waves propagating in an elastic half space covered by an elastic layer (sometimes called 'generalised Lamb waves'). If the covering elastic layer is stiffer than the elastic half space, then it can cause increase in velocity of the Rayleigh-type wave with frequency until it becomes larger than shear wave velocity in the supporting half space, at which point the Rayleigh wave becomes leaky [23].

Note in this connection that Rayleigh waves in a homogeneous elastic half space exist at all frequencies and do not have velocity dispersion. Kramers-Kronig relations are, of course, applicable to such waves. In this case $n(\omega)=1$, and Kramers-Kronig relation (4) gives a correct albeit trivial result $Im\ n(\omega) = 0$, as expected. However, if material losses, e.g. due to relaxation, are taken into account, these relations become of practical importance since in many cases Rayleigh waves are used for physical acoustic investigations of materials. In this case Kramers-Kronig relations for Rayleigh waves are valid, and they are not different from such relations for bulk longitudinal or shear elastic waves propagating in unbounded solids.

### 4.4 Wedge elastic waves

Finally, we will make a few comments on the applicability of Kramers-Kronig relations to wedge elastic waves (also known as wedge acoustic waves), i.e. localised vibration modes propagating along sharp edges of wedge-like elastic structures [23] (see e.g. Fig. 5).

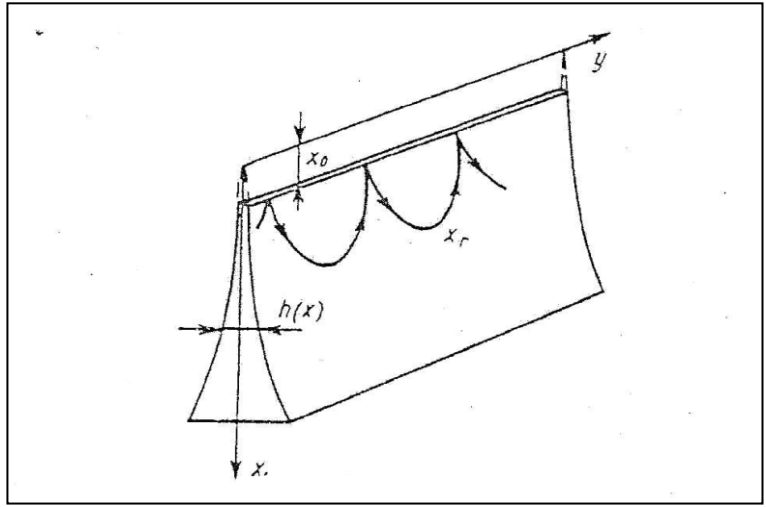

Fig. 5 Truncated elastic wedge of quadratic profile: $h(x) = \varepsilon x^2$; solid lines with arrows illustrate propagation of wedge elastic waves in y-direction as a series of successive reflections of flexural waves from the wedge tip and turnings back due to the total internal reflection [25].

Although such waves are not surface waves in a strict sense (they propagate along a line of intersection of two surfaces, and therefore they are often called 'line waves'), it is convenient to briefly discuss them here since many of their properties are similar to the properties of the above-mentioned surface waves and waves propagating in open waveguides.

Moreover, the theory of wedge elastic waves propagating in slender wedges of arbitrary profile can be constructed using the geometrical acoustics approach in the same way as it is done for normal waves in open waveguides [23, 24].

Like in the case of open acoustic or elastic waveguides discussed above, Kramers-Kronig relations have limited applicability to wedge elastic waves. Similarly to the case of Rayleigh waves in a homogeneous half space, they are applicable to wedge elastic waves propagating in wedges of linear profile (formed by intersection of two planes) as there is no dispersion in this case because of the absence of any linear dimensions (wedges of linear profile are characterised by a single non-dimensional parameter - the wedge angle $\theta$). However, for any other profiles, dispersion of wedge waves does exist, and this may be associated either with the appearance of branch points of functions $n(\omega)$ outside the real frequency axis describing wave energy leakage or with not 'good' behavior of $n(\omega)$ at very large frequencies. In all such cases Kramers-Kronig relations may become not applicable to wedge elastic waves.

Typical examples are a truncated wedge of linear profile [24] and a truncated wedge of quadratic profile [25]. Note that in the latter case (see Fig. 5) the theory is very similar to that for open underwater acoustic waveguides with a linear dependence of sound velocity on depth. The approximate solutions of the resulting dispersion equation show that $n(\omega)$ contains exponential terms [25], which means that Kramers-Kronig relations become not applicable to localised waves in wedges of quadratic profile.

## 5. Conclusions

Kramers-Kronig dispersion relations have limited applicability to normal modes propagating in waveguides. As a rule, they may be applicable to waveguides for which energy leakage to the surrounding media is not possible and the functions of mode refraction coefficient $n(\omega)$ behave well at very large frequencies, e.g. acoustic waveguides with rigid walls. If such waveguides contain ideal media (e.g. ideal liquids), Kramers-Kronig relations provide a general relationship between the velocities of propagating normal modes above the cut off frequencies and the attenuation decrements of non-propagating (evanescent) modes below the corresponding cut off frequencies. If causal material losses in the media, e.g. due to relaxation, are taken into account, Kramers-Kronig relations for such waveguides remain valid and connect the combined dispersion due to relaxation and waveguide propagation with the combined attenuation of propagating and evanescent waves.

If waveguides are open, e.g. formed by layers with different sound velocities, and there is no energy leakage from such waveguides, the applicability of Kramers-Kronig relations to normal modes of such waveguides still can be questionable, depending on 'good' behaviour of the mode refraction coefficient $n(\omega)$ at high frequencies. If energy leakage from open waveguides is possible at a certain frequency range, then Kramers-Kronig relations break down.

The same conclusions can be drawn for surface and wedge elastic waves. If surface or wedge waves become leaky waves at certain frequency ranges or if their velocities do not show 'good' behavior at high frequencies, Kramers-Kronig relations generally become invalid.